\documentstyle[aps,prl,twocolumn]{revtex}

\begin{document}
\title{Maximal Height Scaling of Kinetically Growing Surfaces}
\author{Subhadip Raychaudhuri$^{(1)}$, Michael
Cranston$^{(2)}$, Corry Pryzybla$^{(1)}$, and Yonathan Shapir$^{(1)}$}
\address{$^{(1)}$Department of Physics and Astronomy, University of
Rochester, Rochester, NY 14627\\
$^{(2)}$Department of Mathematics, University of Rochester, Rochester,
NY14627}
\date{\today}
\maketitle
\begin{abstract}
The scaling properties of the maximal height of a growing self-affine
surface with a lateral
extent $L$ are considered. In the late-time
regime its value measured relative to the evolving average
height scales like the roughness: $h^{*}_{L} \sim L^{\alpha}$. 
For large values its distribution obeys $\log{P(h^{*}_{L})} \sim -A({h^{*}_{L}}/L^{\alpha})^{a}$,
charaterized by the exponential-tail exponent $a$.
In the early-time regime where the roughness grows as 
$t^{\beta}$, we find 
$h^{*}_{L} \sim t^{\beta}[\ln{L}-({\beta\over \alpha})\ln{t} + C]^{1/b}$
where either $b=a$ or
$b$ is the corresponding exponent of the velocity distribution.
These properties are derived from scaling and extreme-values arguments.
They are corroborated by numerical simulations and supported by exact
results for surfaces in $1D$ with the asymptotic behavior
of a Brownian path.
\end{abstract}


   Kinetic roughening of self-affine surfaces has
continuously attracted much interest in the last two decades.
It is observed in many systems
such as crystal growth, vapor deposition, molecular-beam
epitaxy, electrochemical processes, bacterial growth, burning fronts, etc.
\cite {family1} \cite{bara} \cite{villain} \cite{dkim}.
    Its investigations mostly dealt
with the scaling properties of the surface roughness. Other characteristics,
which have been studied more recently, include: The distribution of the
width \cite{zia} \cite{plis}, the distribution of the
site-average height velocity in the
1D Kardar-Parisi-Zhang (KPZ) \cite{kardar} model \cite{derrida}, 
the density of the extrema \cite{ziadasSarma},
the persistence \cite{PhysicsToday}, and cycling effects\cite{shapir}.
   In the present work we focus on another important property: The
scaling behavior of the highest point of the surface.
In contrast with all characteristics listed above, the maximal height is not 
a site-average property. Rather, it is an {\it extremal} one which takes 
just one single value for every realization of the surface. 
Since the maximal height is one of the two 
most {\it uncommon} values of the height of the whole surface (the second 
being the lowest height which follows a similar behavior), to obtain 
its scaling behavior and that of its distribution, an inherently different 
approach is required. It blends together the ususal scaling 
methodology to the roughness with extreme-value statistics of rare 
events \cite{gumbel}.

Besides the theoretical challenge involved in rare events statistics of 
variables with intricate spatial and temporal
properties, the maximal height is of a 
paramount importance in technological applications. It is the most
significant property for systems with corroded
surfaces in which the deepest, or the weakest, point determines the
onset of breakdown\cite{galambos}. Another such example are batteries in which
a shortcircuit occurs when the highest point of the metal surface accumulated
on one electrode reaches the opposite one. 

 In addition to the absolute height measured from
the substrate, we define the relative
height as the height measured from the evolving average height of
the surface. The scaling properties of the maximal height (MH) and 
the maximal relative height (MRH) are derived from the combined scaling 
and extreme statistics approaches. They are supported by
numerical simulations and by rigorous results from extremal excursion properties
of Brownian paths. Our main findings are: In the late-time regime 
the MRH scales like the roughness and different properties of its distribution 
are elucidated. In the short-time regime the maximal height grows logarithmically 
slower than the roughness and has a logarithmic dependence on $L$ 
(while the roughness is $L$-independent). The exact form of these 
logarithmic terms depends on the late-time distribution of either the 
velocity or the MRH.

   The surface roughness $w(L,t)$, where $L$ is the
lateral size of the system is defined as
$w^{2}(L,t) = \overline{H(\vec{r},t)^{2}} - \overline{H_{L}(t)}^{2}$.
In this definition
$H(\vec r,t)$ is the surface height.
An overbar denotes average over the sites $\vec{r}$ of
the basal plane of size $L^{D}$ ($D=d-1$). In particular,
$\overline{H_{L}(t)}= L^{-D}\sum_{\vec{r}\in L^{D}}H(\vec{r},t)$ is the average
height of a given surface at time $t$, and $\overline{H_{L}(t)} = vt$
(with $v$ being the growth velocity). We also define the roughness as
$W(L,t) = <w^{2}(L,t)>^{1/2}$, where angular brackets
denote average over the ensemble of surface configurations.
We recall that that it obeys the
following scaling form \cite{family1}\cite{family2}:
$W(L,t) \sim  L^{\alpha}g(L/\xi(t))$, where
$\xi(t) \sim  t^{1/z}$ is the lateral correlation length ($z$ is the
dynamic exponent). For large time
$t > t_{\times}(L) = L^{z}$ it reaches its time-independent value:
$W_{L} \sim L^{\alpha}$,
while for $t < t_{\times}(L)$ it is $L$-independent: $W(t) \sim t^{\beta}$,
where $\beta={\alpha\over z}$
is the growth exponent. Some generic models of surface growth are:
Random deposition (RD), Edwards-Wilkinson (EW),
\cite{edwards}, KPZ \cite{kardar}, DasSarma-Tamborenea (DT)\cite{Dasarma}{\it etc.}

The relative height is given by:
\begin{equation} \label{eq:1}
h_{L}(\vec r, t) = H_{L}(\vec r, t) - \overline{H_{L}(t)}
\end{equation}

The height at a site is the sum of the average height and the relative height.
If either one of them fluctuates much more than the other, it will
make the dominant contribution to the spread of the height 
distribution in general and to that of the MH (denoted by $H^{*}_{L}(t)$) in particular.

   A most important distinction with regard to the MH 
is, therefore, between a first category of systems in which the
velocity $v$ at the large $t$ limit, has a non-trivial distribution $P(v)$
(because the instanteous rate of deposition on the surface is affected 
by its configuration),
and a second category of
systems in which $P(v)$ is trivial because the 
instantenous deposition rate is indepnedent of the surface configuration.
Typical systems of the first category are those in the KPZ universality
class \cite{derrida}.
In the second category are, {\it e.g.}, systems in the EW and DT
universality classes with either a constant or a white-noise deposition
rate. 

The rational for this categorization will become apparent in the 
discussion of the early-time regime which is addressed below following
the the analysis of late-time regime.
In this latter regime,
systems from both categories with $P(v) \ne \delta (v-v_{0})$ 
have their MH distribution dominated by that  
of the average height itself. Thus  
their height distributions
are given essentially by those of their respective velocities (multiplied by the 
time). For such systems the
MRH  $h^{*}(\vec r, t) = H^{*}(\vec r, t) - \overline{H_{L}(t)}$  
describes the subdominant fluctuations
and is of interest as well. 
The MH distribution for some important $2^{nd}$ category systems
in which $v$ is not fluctuating ({\it e.g.}, electrodeposition at a 
constant current) is obviously given by that of the MRH, up to a uniform 
shift by $vt$.

We begin by analyzing the behavior of the MRH
in the late-time regime. In this regime the
probability distribution of the relative height is time-independent.
Consistent with the scaling description, the
roughness $W_{L}$ is the only relevant scale of this distribution.
Thus the height distribution depends only on $u={h\over w_{L}}$, and
$P(u)$ is independent of $L$ (by $P(x)$ throughout we denote the
distribution of $x$,
rather than one particular distribution). Therefore the MRH
distribution should only depend on
$u^{*}=h^{*}_{L}/w_{L}$, and be independent of $L$.   
This is confirmed numerically for the distributions of the 1D EW model 
for different $L$, which are depicetd in Fig. 1.

 Our simulations indicate that this distribution vanishes for $h^{*}_{L}
\rightarrow 0$, faster than exponential. For large arguments
$h^{*}_{L} \rightarrow \infty $, the distribution
also drops faster than exponential
and could be fitted to a behavior
$\sim\exp[\{-A({h^{*}_{L}}/L^{\alpha})^{a}\}]$ with the 
exponent $a=2$. Therefore all 
moments $<(h^{*}_{L}/W_{L})^{k}>$ (with $k$ positive or negative) are finite. 
In particular, both the average and the typical MRH scale like $W_{L}$.

While the MRH distribution for this model depends on the boundary 
conditions (BC), its 
functional decay at large values does not. We find a behaior
$\sim\exp[\{-A({h^{*}_{L}}/L^{\alpha})^{a}\}]$ with $a=2$ 
independent of the BC \cite{mike}.
(in contrast, the amplitude $A$ varies with the BC).
This is not the case for the behavior of $P(h^{*}_{L}/W_{L})$ 
near the origin which is BC sensitive.

  These numerical observations are reinforced by analytical results
for the 1D Brownian path which faithfully describes
the 1D EW and KPZ surfaces in the late-time regime.
We represent them as a Brownian path with fixed BC $H(0)=H(L)$
({\it i.e.} as a Brownian bridge), allowing the average height to fluctuate.
For the statistical properties of $h(r)=H(r)-\overline {H_{L}}$   
of this Markovian process, these BC are equivalent to periodic ones.

The statistical behavior of the maximal height may be reflected either by
the MRH $h^{*}_{L} = \overline{H^{*}_{L} - H}$
or by  $\Delta_{L} = \overline{(H^{*}_{L} - H)^{2}}$.
More complete results are obtained for the distribution of
$\Delta_{L}$. However, $h^{*}_{L}$ and $\Delta_{L}$ are simply related by:
${h^{*}_{L}}^{2} =  \Delta_{L} - w_{L}^{2}$.

Averaging this equation
over different surface realizations yields:

\begin{equation} \label{eq:2}
   <{h^{*}_{L}}^{2}> =  <\Delta_{L}> - {W_{L}}^{2},
\end{equation}

Using path-decomposition techniques \cite{subha},
we find the Laplace transform (generating
function) of the distribution $P(\Delta_{L})$:

\begin{equation} \label{eq:3}
 \hat {P}(\lambda) = <exp\{-\lambda \Delta_{L}\}> =
 \Bigg {[}\frac {\sqrt{2 \lambda  L}} {\sinh(\sqrt{2\lambda
 L})}\Bigg {]}^{3/2}.
\end{equation}

This immediately implies that $LP(\Delta_{L})$
depends only on  $\Delta_{L}/L$. $LP(\Delta_{L})$ behaves as:
$(\frac{L}{\Delta_{L}})^{5/4} \exp (-\frac{9L}{8 \Delta_{L}})
[1 - \frac{5}{8} \sqrt{\frac{\Delta_{L}}{L}}-
\frac{5}{384}\frac{\Delta_{L}}{L}+\ldots] $
for asymptotically small values of $\Delta_{L}/L$,
and as $(\frac{\Delta_{L}}{L})^{1/2}
\exp (-\frac{\pi ^{2} \Delta_{L}}{2L})
[1 + c_{1}(\frac{L}{\Delta_{L}}) + c_{2}(\frac{L}{\Delta_{L}})^{2} +....] $
for large values.

Moreover, we find that $<\Delta_{L}> = L/2$.
Using the result in Ref. \cite{zia}, ${W_{L}}^{2} = L/12$,
we deduce from Eq. (2): $<{h^{*}_{L}}^{2}>  = 5L/12$, confirming
the heuristic arguments above.
Numerically the distributions of $\Delta_{L}$ and of
$({h^{*}_{L}}^{2})$ are
very similar through the whole range and have the same exponential
decay at large values. 

We stipulate that the robustness (with respect to the BC) of the functional decay 
at large values of the MRH 
distribution is a general property of correlated surfaces.
What is the actual functional decay for a given surface, and whether it 
is related to the distributions of other properties (like that of 
the height itself, the width, etc.), are left as open questions for 
future studies.
For most of the surfaces we expect this decay to be exponential, charaterized 
an exponential-tail exponent $a$. We also suggest that $a=2$ might hold for any Gaussian 
surface. Numerical simulations for the DT model \cite{Dasarma} in both 1D 
and 2D and for the KPZ model in 2D are clearly showing the
average MRH to scale as the roughness. The tails of their 
distributions are exponential 
and consistent with the value of $a=2$\cite{subha}.

Having considered the late-time regime we address now the
early-time one. In this regime the correlation length $\xi$ is 
smaller than $L$ and growing with time. Hence the height variables
are uncorrelated if they are separated by a distance $l > \xi$.
So it is beneficial to begin by exploring the random deposition (RD)
model for which the height variables are totally uncorrelated.
We thus consider $h(i)=H(i)-\overline{H_{L}(t)}$, $i=1,2,\ldots,N$ ($N=L^{D}$) 
random uncorrelated
variables. Each of
them has a binomial distribution which, for large time, converges into
a Gaussian distribution of zero mean and variance $W(t) \sim t^{1/2}$.
Extreme-values statistics then imply a scaling of the
average (or typical) MRH, to leading order, as $W(t)\sqrt{2\ln{N}}$
(the distribution of the MRH is the Gumbel distribution \cite{gumbel}).
Since the average height $\overline{H_{L}(t)}$ is essentially uncorrelated 
with each of the $h(i)$, the MH distribution is that of the MRH shifted by 
$\overline{H_{L}(t)}$.

For realistic models with a growing $\xi(t)$, we may imagine
dividing the system into $N_{\xi} = (L/\xi)^{D}$ cells, such that
the surface is correlated within the cells but no correlations
exist between $H(\vec r)$ belonging to different cells. Every cell has
its own highest point, and they are also uncorrelated. Thus, we
aim to apply again the extreme-values statistics of $N_{\xi}$
independent variables in order
to estimate the MH and the MRH of the whole system.
In this regime both have essentially the same distribution up to a trivial
shift by the average height. This results from
the fact that the MH of a cell has very weak (like a 
power of $\xi/L$) correlations
with the average height of the whole surface.

To proceed to find their scaling behavior, we need first to find the 
functional decay of the distribution
of the MH within every cell. To that aim, we have to determine
which of the cell average height or the cell MRH 
is more dominant ({\it i.e.} fluctuates more).
For systems in the $2^{nd}$ category
their respective fluctuations may be estimated. The
fluctuations in the average height
of a cell are of order $\sqrt{t}{\xi}^{-D/2}
\sim \xi^{z{\beta}^{\prime}}$, with ${{\beta}^{\prime}}= (1-D/z)/2$.
They have to be compared with the cell MRH fluctuations which scale as the
cell roughness: $W_{\xi} \sim \xi^{z{\beta}}$.
For linear models ${{\beta}^{\prime}}={\beta}/2$,
and the MRH fluctuations are dominating.
${{\beta}^{\prime}} < {\beta}$
holds for all systems of the $2^{nd}$ category we have checked.
We conclude that for such systems the fluctuations in the average height
may be discarded and therefore the decay of cell MH distribution
is essentially that of the cell's MRH.

Next we need to find the large-value decay   
of the distribution of the cell MRH.
We expect it to follow that of
$P(h^{*}_{L}/W_{L})$ in the late-time regime,
with $L = \xi$. $P(h^{*}_{L}/W_{L})$, however,
is only defined for specific BC.
Trying to assign such specific BC between the virtual cells
is, of course,  meaningless. Nevertheless, we
will rely on the property, discusses above, that for {\it all BC this
distribution has the same functional decay for large
values of its argument}.  Therefore, 
this universal decay with respect to the BC 
persists even when the BC are ill-defined.
It follows that $P(h^{*}_{\xi})$ of a single cell MRH is most likely 
to decay as $\sim\exp[\{-A({h^{*}_{\xi}}/{W_{\xi}})^{a}\}]$, and so does the
distribution of the MH within every cell (recall that the cell MH and MRH 
differ by a weakly-fluctuating average height).  

Extremal-values statistics then predict the highest of the maximal
heights of $N_{\xi}$ cells to scale as:
$H^{*}_{L} \sim W_{\xi}(\ln{N_{\xi}})^{1/a}$.
Replacing $W_{\xi}$ by $Bt^{\beta}$ and $N_{\xi}$ by
$[L/t^{\beta/\alpha}]^{D}$ we obtain (after shifting by the 
uncorrelated
$\overline{H_{L}(t)}$):
\begin{equation} \label{eq:4}
<h^{*}_{L}> = K t^{\beta}[\ln{L}-({\beta\over \alpha})\ln{t} + C]^{1/a}
\end{equation}
with non-universal constants K and C.

Two conclusions follow:
{\it (i)} The growth of the RMH is slower than that of the
roughness by a logarithmic factor. This is seen in the simulation
results in Fig. 2
where we have plotted $(h^{*}_{L}/W_{L})^{2}$ vs
$\log t$ for the
1D EW model. The plot is consistent with the leading
linear behavior predicted by Eq. (4). The deviation from linearity is in
the direction expected from the next order
correction ($\sim \log \log N_{\xi}$  \cite{gumbel}) which becomes increasingly
important near the crossover to the late-time regime (as $N_{\xi}$ becomes smaller).
{\it (ii)} The MRH depends logarithmically on the system size $L$ 
for $t<t_{\times}$ (when the roughness is $L$-independent).
This may be observed in Fig. 3 for the same model.
Early-time simulations for the DT model 1D and 2D yield
a similar behavior \cite{subha}.

For systems in the first category the early-time behavior might
differ. In such systems the possibility of ${{\beta}^{\prime}} > {\beta}$ 
within a cell cannot be ruled out (in such a case the growth exponent of the 
surface assumes the value of ${{\beta}^{\prime}}$).

This is scenario onserved in the 1D KPZ model 
for which
the width of the surface relative height distribution ${W_{\xi}}$ was 
calculated numerically. It was found to
behave as  $\exp[-(h/{W_{\xi}})^{\eta}]$, with $\eta=2$ at the center of
the distribution, $\eta_{-} \simeq 1.6$ in the left tail and $\eta_{+}
\simeq 2.4$ for the right tail of largest heights \cite{bray}.
This behavior
mimics very closely that of the velocity distribution in the
late-time regime (with $\eta_{-}=1.5$ and $\eta_{+}=2.5$, 
respectively \cite{derrida}).
Thus, the early-time extreme values of the height are
determined by {\it the fluctuations in the average height of a cell, and
not from these of the cell MRH} (in contrast with second category systems
discussed above).
   Therefore, the power $1/a$ of the logarithmic term in Eq. (4) has
to be replaced by $1/\eta_{+}$, at least for $\xi << L$. In this regime
corrections to scaling might still be important while for larger values
of $\xi$, $N_{\xi}$ becomes smaller and the largest typical height moves
towards the center of the distribution with possibly an effective exponent
$\eta$ between $2.5$ and $2$.
Similar arguments were used recently \cite{duxbury}
to explain the extremal statistics in the energetics of
random-bond interfaces.

In summary we have obtained the scaling behavior of the
maximal height. In the short-time regime its
growth differs from that of the roughness by a logarithmic correction 
depending on $t$ and $L$. 
The exponent of this correction is determined from the exponential decay of the
distribution of either the velocity ($1^{st}$ category) , or that of the
MRH ($2^{nd}$ category), in the late-time
regime. In this latter
regime the surface is correlated throughout the whole system and the
average MRH scales like the roughness.
It will be most welcome if more analytical and numerical results
for the MH and the MRH (and their distributions)
of other models could be obtained. 
Finally, we hope that experimental systems with accurate
roughness scaling over a wide range will exhibit the scaling behavior
of the maximal height as well.

We acknowledge fruitful communications
with D. Foster, J. Jorne, T.- M. Lu, and G.- C. Wang on their
experiments. This work was supported by the ONR grant N00014-00-1-0057 (SR, YS),
and by the NSF grants DMS-9972961 (MC) \& REU-9987413 (CP).

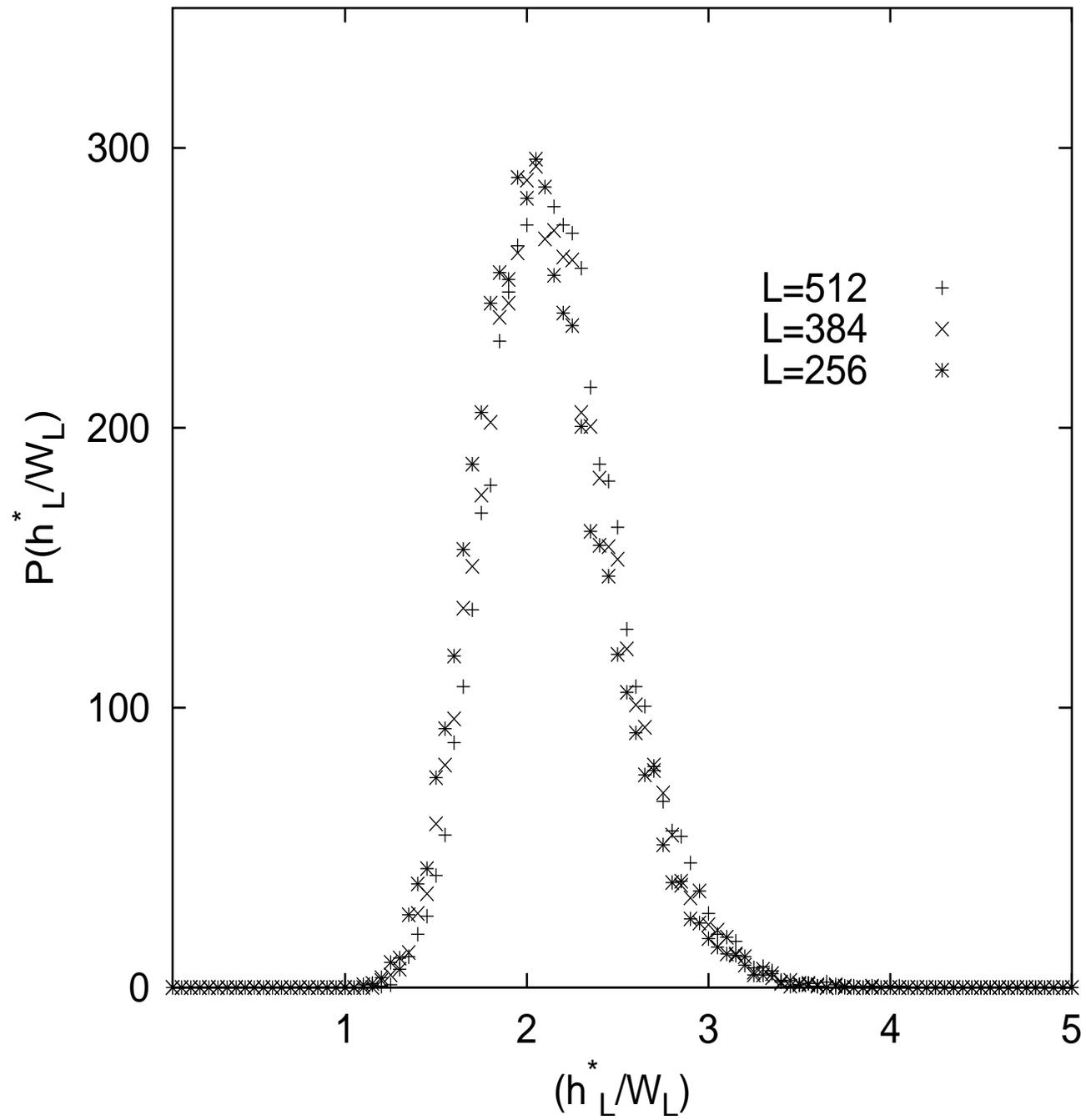
\begin{figure}
\caption{The distribution of $h^{*}_{L}/W_{L}$ of the 1D EW model with
periodic BC for different $L$. All lengths are expressed in terms of the lattice spacing.}
\label{Fig. 1}
\end{figure}

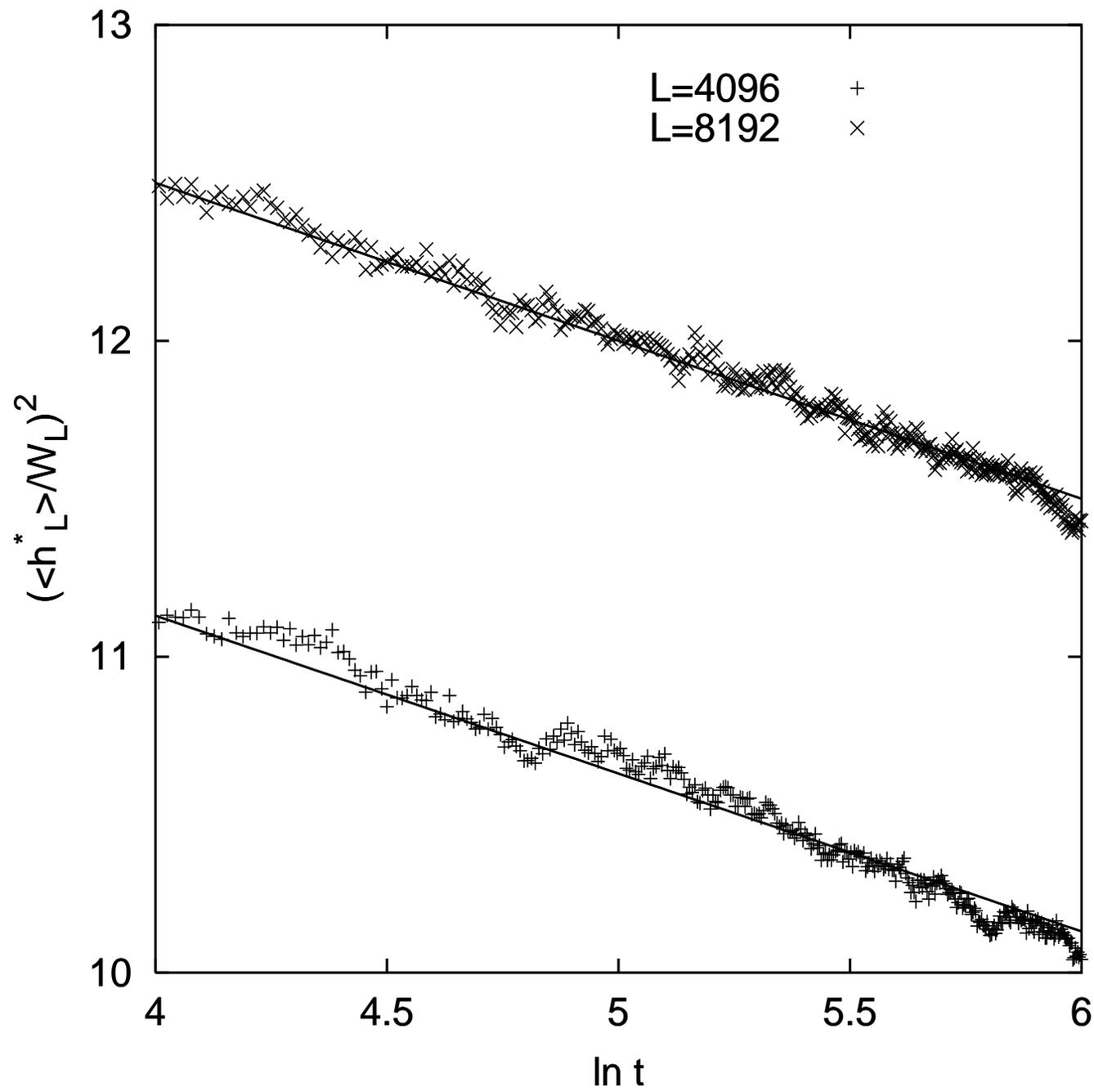
\begin{figure}
\caption{ $(h^{*}_{L}(t)/W_{L}(t))^{2}$ $vs$ $\log t$ of the 1D EW
model for two values of $L$. The straight lines are guide for the eyes.}
 \label{Fig. 2}
\end{figure}

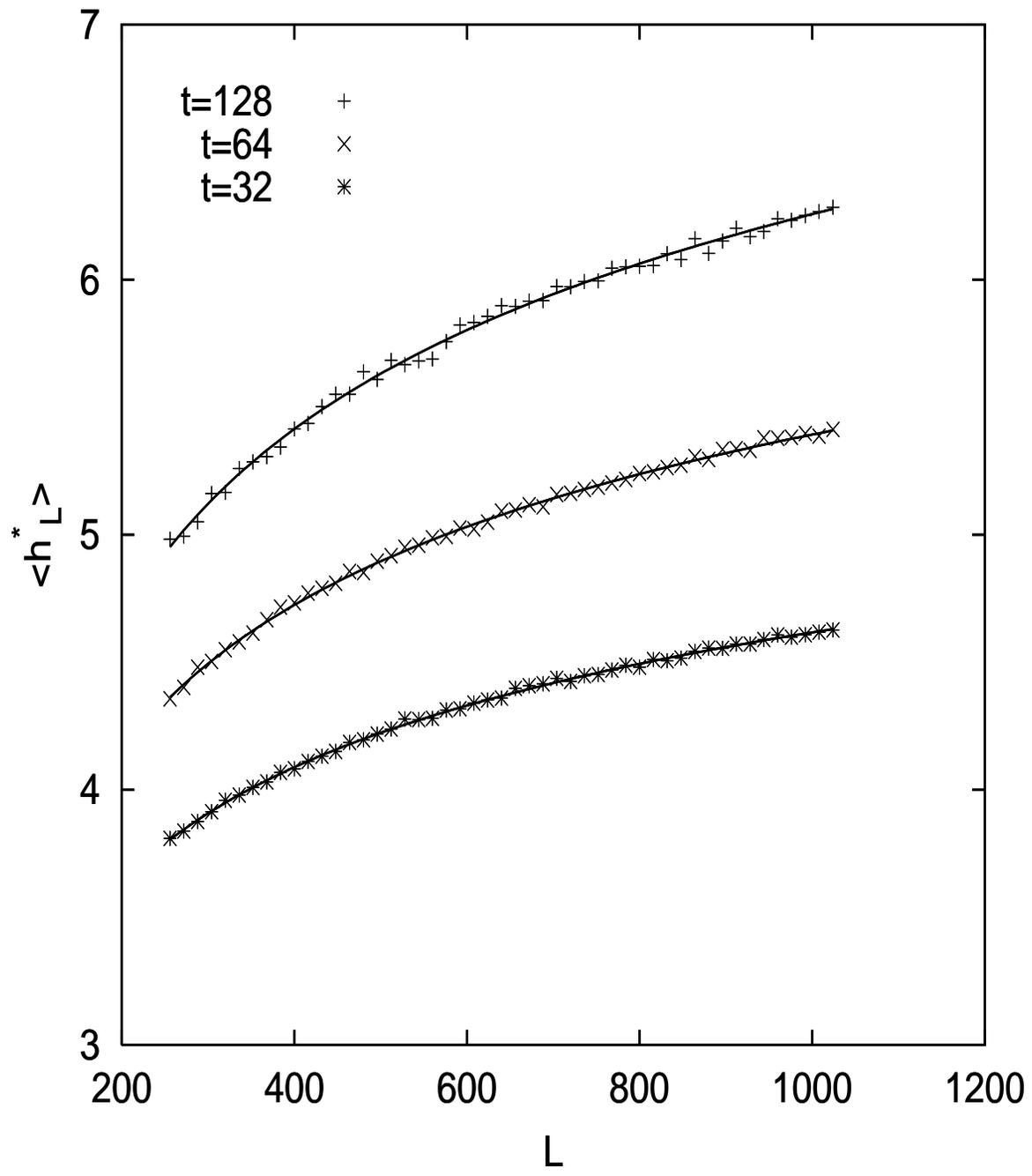
\begin{figure}
\caption{ $h^{*}_{L}(t)$ {\it vs} $L$ of the 1D EW model for different 
values of $t$. The solid lines are fits to ${\bar K}t^{\beta}[\log L + {\bar C}(t)]^{1/2}$.}
\label{Fig. 3}
\end{figure}


\begin{references}

\bibitem{family1}
{\it Dynamics of Fractals Surfaces}, edited by F. Family and T.
Vicsek (Cambridge University Press, Cambridge, England, 1990).

\bibitem{bara}
A.-L. Barabasi and H. E. Stanley, {\it Fractal Concepts in Surfaces
Growth}, (Cambridge University Press, Cambridge, England, 1995).

\bibitem{villain}
J. Villain and A. Pimpinelli, {\it Physique de la Croissance
Crystalline}, (Editions Eyrolles, Paris, France, 1995).

\bibitem{dkim}
{\it Dynamics of Fluctuating Interfaces and Related Phenomena},
edited by D. Kim, H. Park, and B. Khang (World Scientific, Singapore,
1997).

\bibitem{zia}
G. Foltin, K. Oerding, Z. Racz, R. L. Workman, and R. K. P. Zia,
Phys. Rev. E {\bf 50}, R639, (1994).

\bibitem{plis}
Zoltan Racz and Michael Plischke, Phys. Rev. E {\bf 50}, 3530, (1994).

\bibitem{kardar}
M. Kardar, G. Parisi, and Y.-C. Zhang, Phys. Rev. Lett. {\bf 56},
889 (1986).

\bibitem{derrida}
B. Derrida and J. L. Lebowitz, Phys. Rev. Lett. {\bf 80}, 209, 1998;
B. Derrida and C. Apert, J. of Stat. Phys. {\bf 94}, 1, 1999.

\bibitem{ziadasSarma}
 Z. Toroczkai, G. Korniss, S. Das Sarma, and R. K. P. Zia,
Phys. Rev. E {\bf 62}, 276, (2000).

\bibitem{PhysicsToday}
Z. Toroczkai and E. D. Williams, Physics Today, 24, December (1999);
J. Krug et al. Phys. Rev. E {\bf 56}, 2702, (1997); H. Kallabis and
J. Krug, cond-mat 9809241; S. N. Majumdar, Curr. Sci. {\bf77}, 370
(1999).

\bibitem {shapir}
Y. Shapir, S. Raychaudhuri, D.G. Foster, and J. Jorne, Phys. Rev. Lett.
84, 3029 (2000).

\bibitem{gumbel}
E. J. Gumbel, {\it Statistics of Extremes}, (Columbia University Press,
NY, 1958).

\bibitem{galambos} 
See, {\it e.g.}, Chapter I in {\it Extreme Value Theory 
and Applications}, p. 15, edited by J. Galambos, J. Lechner, and E. Simiu
(Kluwer Academic Publishers, Dordrecht, Netherlands, 1994). 

\bibitem{edwards}
S. F. Edwards and D. R. Wilkinson, Proc. Roy. Soc. London A {\bf 381},
17 (1982).

\bibitem{family2}
 F. Family and T. Vicsek, J. Phys. A{\bf 18}, L75 (1985).

\bibitem{mike} The same large value exponent holds for the
distribution of the range of the surface, defined as
the difference between the maximal and the minimal heights.
K. L. Chung, Acta. Math. Hungar.,  {\bf 39}, 65 (1982).

\bibitem {subha}
S. Raychaudhuri, M. Cranston, and Y. Shapir, to be published.

\bibitem{Dasarma}
S. Das Sarma and P. Tamborenea, Phys. Rev. Lett. {\bf 66}, 325 (1990);
D. E. Wolf and J. Villain, Europhy. Lett. {\bf 13}, 389 (1990).

\bibitem{bray}
J. M. Kim, M. A. Moore, and A. J. Bray, Phys. Rev. A {\bf 44},
2345 (1991).

\bibitem{duxbury}
E. T. Seppala, M. J. Alava, and P. M. Duxbury, cond-mat/0102318.

\end{references}
\end{document}